\documentclass[epjCONF]{svjour}

\usepackage{epsfig}     %\  Used to include figures
\usepackage{graphics}
\usepackage[varg]{txfonts} % Times fonts
\usepackage[latin1]{inputenc}
\session-title{Conference Title, to be filled}
\def \xmm {\textit{XMM-Newton}}
\newcommand{\bc}{\begin{center}}
\newcommand{\ec}{\end{center}}
\def\ltsima{$\; \buildrel < \over \sim \;$}
\def\lsim{\lower.5ex\hbox{\ltsima}}
\def\gtsima{$\; \buildrel > \over \sim \;$}
\def\gsim{\lower.5ex\hbox{\gtsima}}
\def\lsun{~L_{\odot}}
\def\msun{~M_{\odot}}
\def\rsun{~R_{\odot}}
\def\zsun{~Z_{\odot}}
\def\mdot {\dot M}

\def\hd {HD\,49798}

\def\bd {BD\,+37$^\circ$\,442}
\begin{document}
\title{X-ray emission from hot subdwarfs with compact companions}
\author{Sandro  Mereghetti\inst{1}\fnmsep\thanks{\email{sandro@iasf-milano.inaf.it}} \and Nicola La Palombara\inst{1} \and Paolo  Esposito\inst{1} \and
Andrea Tiengo\inst{2,1}}
\institute{INAF - IASF Milano, via Bassini 15, I--20133 Milano, Italy \and
IUSS, piazza della Vittoria 15, I--27100 Pavia, Italy}
\abstract{
We review the X-ray observations of hot subdwarf stars.
While no X-ray emission has been detected yet from binaries containing B-type subdwarfs, interesting
results have been obtained in the case of the two luminous O-type subdwarfs \hd\ and \bd  .
Both of them are members of binary systems in which the X-ray luminosity is powered by
accretion onto a compact object:  a rapidly spinning (13.2 s) and  massive (1.28 $\msun$) white dwarf in the case of \hd\ and
most likely a neutron star, spinning at 19.2 s, in the case of \bd .
Their study can shed light on the poorly known processes taking place during common envelope evolutionary
phases and on the properties of wind mass loss from hot subdwarfs.
} %end of abstract
\maketitle
\section{Introduction}
\label{intro}

Hot subdwarfs are evolved low--mass stars that lost most of their hydrogen envelopes and are now in the stage of helium core burning.
A possible mechanism responsible for the loss of the massive H envelopes necessary to form hot subdwarfs is
mass transfer in a binary.
Indeed, many hot subdwarfs are members of binary systems \cite{max01,mor03},
supporting the idea that non-conservative mass transfer played a role in the formation of these stars.
Evolutionary models  predict that the most of the subdwarf  companions  in binaries with orbital period shorter than $\sim$10 days
should  fall into two main groups: late type main sequence stars and white dwarfs \cite{han02}.

X-rays can be used to identify hot subdwarfs with compact companions: they can originate from surface
thermal emission of neutron stars or sufficiently hot white
dwarfs or can be produced if the compact object accretes mass from the subdwarf at a sufficiently high rate.
This is well illustrated by our recent results on the sdO binaries discussed below: \hd\ and \bd .
Furthermore, accreting  compact objects can be used  as probes to investigate the  poorly known properties of
the stellar winds of hot subdwarfs.

\section{A massive white dwarf with a sdO companion}
\label{sec:3}

The peculiar properties of the bright blue star \hd\ (B=8, B$-$V=$-$0.27)  attracted the attention of many astronomers since the sixties,
when it was included   in the (then small) group of early type subdwarfs \cite{jas63}.
Its first spectroscopic observations  showed a dominance of He and N lines and
radial velocity variations, pointing to a binary nature, that was later confirmed with the
discovery of  the orbital period of 1.5477 days \cite{tha70}.
In the following years several studies concentrated on a detailed modeling of the star's
atmosphere. \hd\ was classified as a subdwarf of spectral type O6,  with  effective temperature
T$_{eff}$ = 47,500 K, and surface gravity log \textsf{g} = 4.25$\pm$0.2 \cite{kud78}.
The overabundance of   He and N,
%equalling H in number (X$_{H}$=0.19 and X$_{He}$=0.78),
and the low abundances of C and O, indicated that \hd\ is the stripped core of an initially much more massive and larger star.
The optical mass function could be measured with great precision
(f$_{OPT}$=0.263$\pm$0.004 $\msun$, \cite{sti94}), but all the attempts to reveal the companion star,
outshined  in the optical/UV by the very luminous sdO (L$\sim$10$^4$ $\lsun$), were unsuccessful.

The nature of the "invisible" companion of \hd\ was partially clarified only in 1996, thanks to the
\textit{ROSAT} discovery of soft X-rays showing periodic pulsations at 13 s \cite{isr97}. This clearly
pointed to the presence of a collapsed object. However,  with the poorly constrained spectrum obtained with \textit{ROSAT},
it was impossible to distinguish between a neutron star or a white dwarf.
In fact, depending on the spectral assumptions, the derived  X-ray luminosity resulted
in the range from $\sim$10$^{32}$ erg s$^{-1}$ up to $\gsim$10$^{35}$ erg s$^{-1}$,
compatible with both possibilities.

\begin{figure}[h]
\centering
\includegraphics[width=9cm,angle=0]{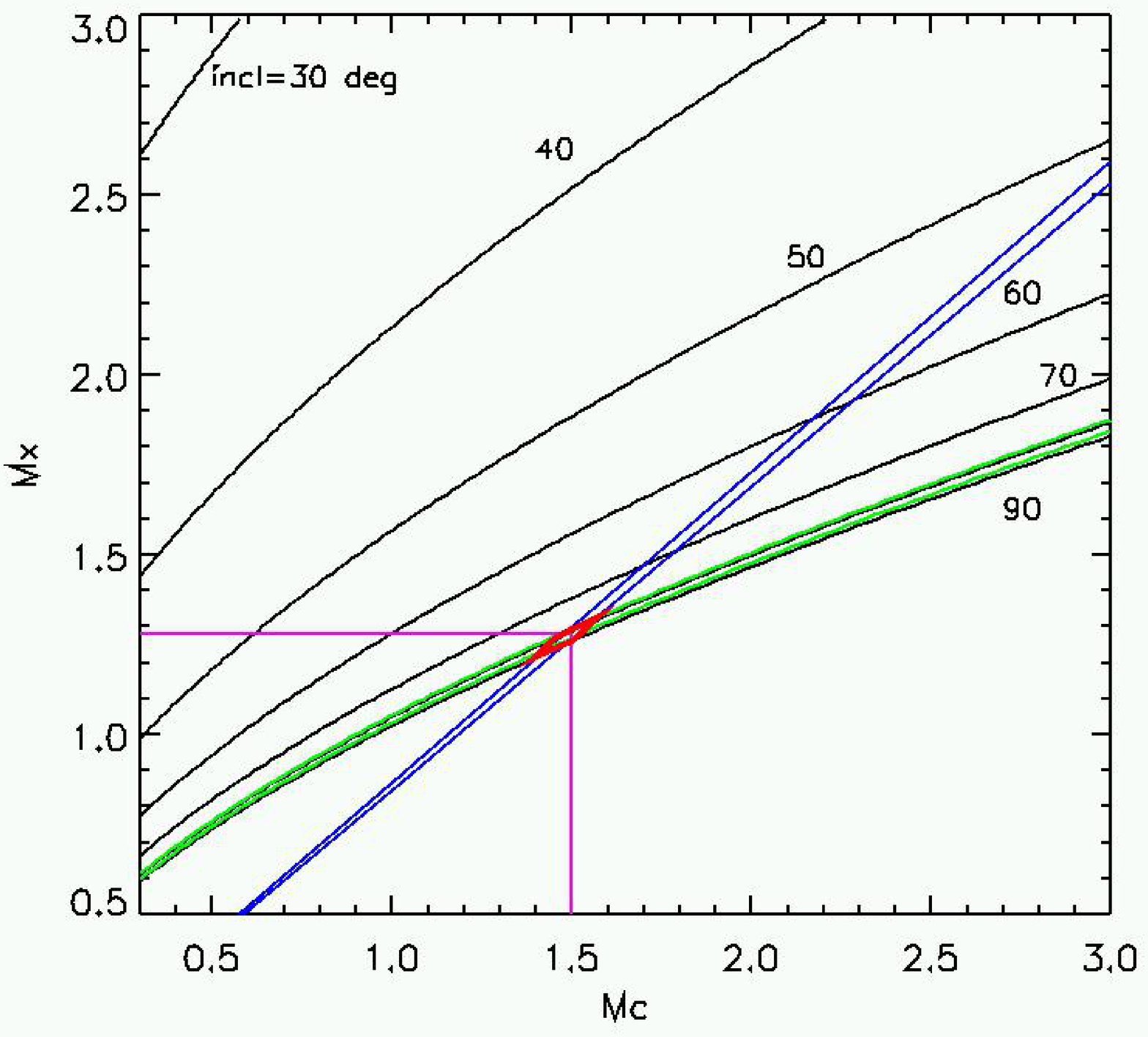}
\caption{Masses of HD 49798 (\textsl{X axis}) and of its white dwarf companion (\textsl{Y axis}) in units of solar masses.
 The curved lines give, for different inclination
 angles, the constraints from the optical mass function. The straight lines indicate  the mass ratio interval derived combining
 the X-ray and optical mass function. The eclipse duration constrains the inclination in the range 79-84$^\circ$.}
\label{limits}
\end{figure}

To further advance in the understanding of this system we had to await for a long \xmm\ pointing carried out in 2008.
We strategically scheduled this observation at the orbital phase of the expected X-ray eclipse, that
was never covered in previous X-ray observations.  Our main
objectives were to exploit the regular X-ray periodicity, which makes this system equivalent to a double spectroscopic binary,
to constrain the masses of the two stars, and to get a better estimate of  the source spectral parameters and luminosity.
The measurement of the X-ray pulse delays induced by the orbital motion, together with the discovery of an X-ray eclipse lasting $\sim$1.3
hours, allowed us to derive the X-ray mass function as well as the system's inclination.
This information, coupled to the already known optical mass function, gives the masses of the two stars: M$_{sd}$ =
1.50$\pm$0.05 $\msun$ for the subdwarf \ and M$_{WD}$ = 1.28$\pm$0.05 $\msun$ for its companion (Fig.\ref{limits}).
Furthermore, the high quality spectrum obtained with  \xmm\ instrument showed that the total luminosity is only
$\sim$10$^{32}$ erg s$^{-1}$, much smaller than that expected from a neutron star accreting in the stellar wind of \hd .
We thus concluded that this system most likely contains one of the most massive white dwarfs with a dynamical mass
measurement, which is also the one with the shortest spin period (P=13 s) \cite{mer09}.

\begin{figure}[h]
\centering
\includegraphics[width=14.5cm,angle=0]{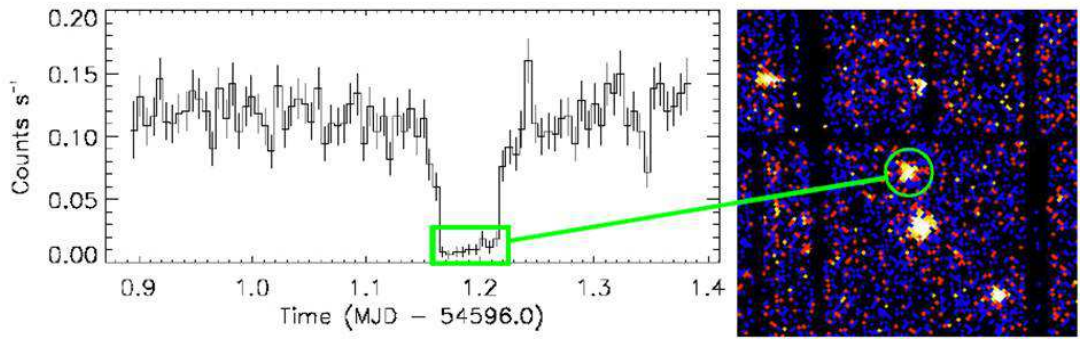}
\vspace{-4cm}
\caption{ X-ray light curve of the May 2008 observation of \hd . An eclipse lasting 1.3 hours is clearly visible. The image accumulated
during the eclipse time interval shows the presence of significant emission in the 0.2--10 keV band.}
\label{ecl}
\end{figure}

The radius of \hd\ (1.45$\pm$0.25 $\rsun$) is much smaller than
that of its Roche-lobe, so accretion onto the white dwarf must occur through stellar wind capture.
In fact, \hd\  is one of the few O-type subdwarfs  for which evidence for a relatively strong
stellar wind has been obtained from optical/UV spectroscopy. Modelling of its P-Cygni line profiles yields
a  mass loss rate of $\sim$3$\times$10$^{-9}$  $\msun$ yr$^{-1}$   \cite{ham10}.

The X-ray emission observed in the \hd\ binary consists of a very soft
and strongly pulsed blackbody-like component   (kT=40 eV), plus  a harder spectral component that
dominates above $\sim$1 keV.  The latter can be fit equally well by a power law with photon index $\Gamma$=1.6 or by a
thermal bremsstrahlung with kT=8 keV, and shows a double peaked pulse profile \cite{mer11b}. Such characteristics are quite
similar to those of cataclysmic variables of the polar and intermediate polar class, despite these systems are very
different for what concerns their mass-donor stars and accretion geometry.

At the end of the current He-burning phase, \hd\ will expand again and transfer He-rich matter through Roche-lobe overflow
during a semi-detached phase \cite{ibe94}, but the fraction of mass that is  retained on the white dwarf is rather uncertain.
Recent  computations, performed assuming the mass accumulation efficiency  that takes into account the wind mass loss
triggered by the He-shell flashes  \cite{kat04}, indicate that a mass of 1.4 $\msun$ can be reached after only a few 10$^4$
years \cite{wan10}. However, there are other critical factors which influence the final fate of the white dwarf, such as,
e.g.,  its composition and rotational velocity.

If \hd\ hosts a CO white dwarf, it could be the progenitor of an over-luminous type Ia supernova, since the fast rotation can
increase the mass stability limit above the value for non-rotating stars. Massive white dwarfs are expected to have
an ONe composition, but again the high spin might play a role here, since it can lead to the formation of  CO white dwarfs
even for high masses.
% \cite{dom96}.
The fact that this system originated from a pair of relatively massive stars ($\sim$8--9 $\msun$) might suggest that it could
be the  progenitor of a type Ia supernova with a short delay time. However, the delay time might be considerably
longer if the explosion has to await that the white dwarf spins down \cite{dis11}.

Alternatively,  if the companion of \hd\ is an ONe white dwarf, an accretion-induced collapse might occur,
leading to the formation of a neutron star. The high spin rate and low magnetic field make
this white dwarf an ideal progenitor of a millisecond pulsar.
The evolution of systems like this one could  be a promising scenario for the direct formation of  millisecond pulsars, i.e.
one not involving the recycling of old pulsars in accreting low-mass X-ray binaries.

Besides their obvious interest in the context of the binary evolutionary modes,
these  X-ray observations are interesting for the study of the sdO star itself. It is remarkable that X-rays
have been detected  also when the white dwarf is eclipsed by the much larger sdO companion (Fig. \ref{ecl}).
The observed emission, with a luminosity of $\sim2\times10^{30}$ erg s$^{-1}$,
could be the first detection of an sdO stars in the X-ray band  \cite{mer11b}.
The observed ratio of X-ray to bolometric luminosity of a few 10$^{-7}$ is in accord to that of  main sequence and supergiant OB stars.
Alternatively, the X-rays seen during the eclipse could be due to the reprocessing of the white dwarf emission in the stellar wind.
Our most recent \xmm\ observations  showing the presence of emission lines of N and Ne
in the eclipse spectrum seem to favor this interpretation.

\begin{figure}[h]
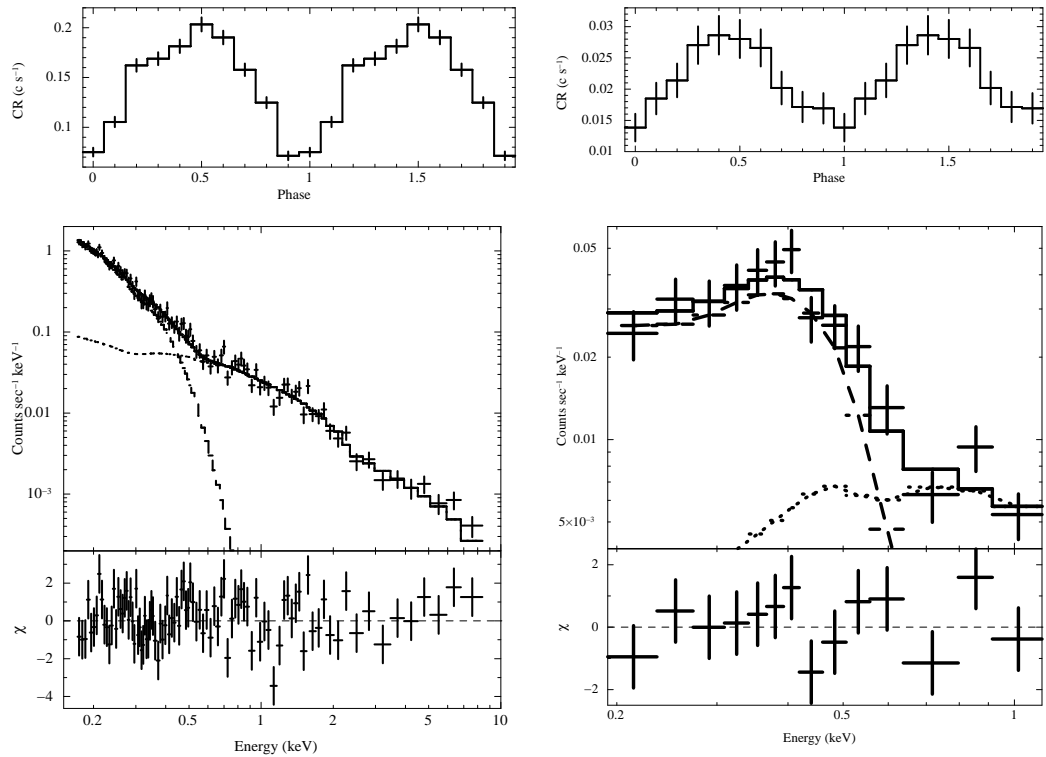

\centering
\resizebox{\hsize}{!}{
\begin{tabular}{c@{\hspace{1pc}}c}
\includegraphics[height=6.1cm,angle=-90]{lc_HD.ps} &
\includegraphics[height=5.9cm,angle=-90]{lc_BD.ps} \\
\end{tabular}
}
\resizebox{\hsize}{!}{
\begin{tabular}{c@{\hspace{1pc}}c}
\includegraphics[height=6.1cm,angle=-90]{spettro_HD.ps} &
\includegraphics[height=5.9cm,angle=-90]{Spettro_BD_new1.ps} \\
\end{tabular}
}
\caption{Comparison of the X-ray light curves and spectra of \hd\ (left) and \bd\ (right).
The top panels show the   light curves in the range 0.2-0.5 keV folded at the periods of 13.2 s and 19.2 s, respectively
for \hd\ and \bd . The bottom panels show the spectra with the best fit blackbody plus power-law models. Note that the flux
of \bd\ is much lower than that of \hd . Hence its spectral parameters are not well constrained. This implies a large uncertainty
on its total X-ray luminosity.
}
\label{figsepc}
\end{figure}

\section{An extreme He star with a neutron star or white dwarf companion}
\label{sec:4}

Prompted by the possible detection of X-ray emission from \hd\ itself during  the white dwarf eclipse,
we requested \xmm\ observations of a supposedly single hot  subdwarf. This led to the
discovery of soft X--rays with a periodicity of 19.2 s from  the luminous sdO   \bd ,
indicating also in this case the presence of  a compact object \cite{lap12}.
This was quite surprising because all the data reported in the literature for  \bd\ did not show any evidence of a binary companion:
infrared observations did not show any excess emission \cite{the95},
and no signatures of a binary nature were seen in  spectroscopic or photometric data \cite{fay73,dwo82,lan73}.
On the other hand,   the   few   radial  velocity measurements  found in the literature give inconsistent values
(V$_r$=--156.4$\pm$1.1 km s$^{-1}$ \cite{reb66},
 V$_r$=--94$\pm$1 km s$^{-1}$ %(Drilling \& Heber 1987),
\cite{dri87}),
which could be caused by binary motion. Further spectroscopic observations are clearly needed to search for
an orbital period in \bd .

The X-ray emission from   \bd\ has a soft spectrum well described by a blackbody with temperature $\sim$45 eV
plus a weak power-law component. This is very   similar to the spectrum  of \hd\ (Fig. \ref{figsepc}).
The best fit X-ray luminosity is $\sim10^{33}$ erg s$^{-1}$ (for a distance of 2 kpc,  \cite{bau95}),
but, due to the uncertainties in the X-ray blackbody temperature, the bolometric X-ray luminosity
could be between $\sim10^{32}$  and $\sim10^{35}$ erg s$^{-1}$.

The observed pulsations clearly indicate the presence of a white dwarf or a neutron star,
most likely powered by accretion from the stellar wind of \bd , which is loosing
mass at a rate of $\sim$3$\times$10$^{-9}$ $M_{\odot}$   yr$^{-1}$ and with
a wind terminal velocity of v$_{\infty}$=2,000\,km\,s$^{-1}$  \cite{jef10}.
Simple computations, assuming a canonical velocity law for the wind and Bondi-Hoyle accretion,
show that a  neutron star orbiting \bd\  with a period of a few days would accrete from the wind
at a sufficiently high rate to produce the observed X-rays. Alternatively, the
accreting companion could be a white dwarf, but this would require  a larger accretion rate that would only be possible if the accretion
is due to Roche-lobe overflow.\\

\begin{figure}[h]
\centering
\includegraphics[width=10cm,angle=0]{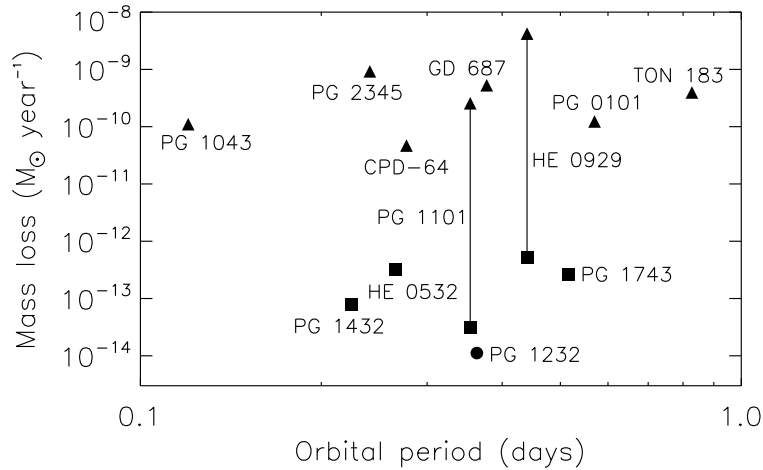}
%\vspace{-5cm}
\caption{Upper limits (3$\sigma$) on the sdB mass loss rates in the case of a white dwarf (triangles), a
neutron star (squares), or a black hole (circle) companion.}
\label{limiti}
\end{figure}

\section{sdB with candidate compact companions}
\label{sec:2}
%and \cite{RefJ}

We observed with the \textit{Swift} satellite a few sdB binaries selected from the MUCHFUSS project,
which aims at finding hot subdwarfs with compact companions  by means of radial and rotational velocity measurements in the optical band
\cite{gei11}.
None of the observed targets was detected in the X-ray band, and we could set upper limits on their luminosity in the range
L$_X\sim10^{30}-10^{31}$ erg s$^{-1}$ \cite{mer11a}.

Although this negative result
does  not allow us to confirm the presence of compact stars inferred from the optical data, the  upper limits on  L$_X$ can be used to
constrain the poorly known properties of the sdB stellar
winds\footnote{We assume that compact objects are really present in these systems. An alternative trivial
explanation for their non-detection is that some of the
assumptions made to infer the presence of compact objects is
wrong. However this is rather
unlikely \cite{gei10} and we will not consider further this possibility.}.

The luminosity upper limits can be converted into limits on the mass loss rates $\mdot_\mathrm{W}$ from the sdB stars,
assuming for simplicity Bondi-Hoyle accretion from the stellar wind onto the compact objects.
This is justified by the fact that in these systems the sdB stars do not fill their Roche-lobes.
Our assumptions lead to conservative upper limits on $\mdot_\mathrm{W}$, since most wind-accreting
neutron stars in high-mass X-ray binaries show an X-ray
luminosity larger than that predicted by the simple Bondi-Hoyle theory.

The resulting limits on $\mdot_\mathrm{W}$ are plotted in Fig.\ref{limiti}, where different symbols are used to
indicate the assumed compact object.
For the systems likely hosting white dwarfs (triangles) the limits are above
$\mdot_\mathrm{W}\sim5\times10^{-11} \msun$ yr$^{-1}$. Although
they are not particularly constraining for the sdB wind properties, we note that they represent one of the few
observational results in this field.
Deeper X-ray observations of the closest candidates (e.g. PG 0101+039 and CPD -64 481) with more sensitive satellites
like \xmm\ or \textit{Chandra} should be able to detect accreting white dwarfs, if their sdB companions lose mass at a
rate $\mdot_\mathrm{W}\sim10^{-12}-10^{-11} \msun$ yr$^{-1}$,
as predicted by theoretical wind models.

More interesting constraints can be inferred from the binaries
likely containing neutron stars or black holes (PG 1432+159, HE 0532-4503, PG 1232-136, and PG 1743+477).
The lack of X-ray emission implies that the sdB stars in these systems have rather weak winds, with
$\mdot_\mathrm{W}<3\times10^{-13} \msun$ yr$^{-1}$, which is
significantly below the predictions of theoretical models \cite{vin02}, if a solar metallicity is assumed.
A metallicity $Z=0.3 \zsun$, or lower, is required for these sdB stars to be consistent with the derived upper limits.\\

\textbf{Acknowledgements
}
We acknowledge financial contribution from the agreement
ASI-INAF I/032 /10/0.

\end{document}